\documentclass[preprint]{aastex631}

\usepackage{ulem}

\submitjournal{ApJL}

\shorttitle{DSA-110 FRB RMs}
\shortauthors{Sherman et al.}

\begin{document}

\title{Deep Synoptic Array Science: Implications of Faraday Rotation Measures of Localized Fast Radio Bursts}

\correspondingauthor{Myles B. Sherman}
\email{msherman@caltech.edu}

\author{Myles B. Sherman}
\affiliation{Cahill Center for Astronomy and Astrophysics, MC 249-17 California Institute of Technology, Pasadena CA 91125, USA.}

\author{Liam Connor}
\affiliation{Cahill Center for Astronomy and Astrophysics, MC 249-17 California Institute of Technology, Pasadena CA 91125, USA.}

\author{Vikram Ravi}
\affiliation{Cahill Center for Astronomy and Astrophysics, MC 249-17 California Institute of Technology, Pasadena CA 91125, USA.}
\affiliation{Owens Valley Radio Observatory, California Institute of Technology, Big Pine CA 93513, USA.}

\author{Casey Law}
\affiliation{Cahill Center for Astronomy and Astrophysics, MC 249-17 California Institute of Technology, Pasadena CA 91125, USA.}
\affiliation{Owens Valley Radio Observatory, California Institute of Technology, Big Pine CA 93513, USA.}

\author{Ge Chen}
\affiliation{Cahill Center for Astronomy and Astrophysics, MC 249-17 California Institute of Technology, Pasadena CA 91125, USA.}

\author{Kritti Sharma}
\affiliation{Cahill Center for Astronomy and Astrophysics, MC 249-17 California Institute of Technology, Pasadena CA 91125, USA.}

\author{Morgan Catha}
\affiliation{Owens Valley Radio Observatory, California Institute of Technology, Big Pine CA 93513, USA.}

\author{Jakob T. Faber}
\affiliation{Cahill Center for Astronomy and Astrophysics, MC 249-17 California Institute of Technology, Pasadena CA 91125, USA.}

\author{Gregg Hallinan}
\affiliation{Cahill Center for Astronomy and Astrophysics, MC 249-17 California Institute of Technology, Pasadena CA 91125, USA.}
\affiliation{Owens Valley Radio Observatory, California Institute of Technology, Big Pine CA 93513, USA.}

\author{Charlie Harnach}
\affiliation{Owens Valley Radio Observatory, California Institute of Technology, Big Pine CA 93513, USA.}

\author{Greg Hellbourg}
\affiliation{Cahill Center for Astronomy and Astrophysics, MC 249-17 California Institute of Technology, Pasadena CA 91125, USA.}
\affiliation{Owens Valley Radio Observatory, California Institute of Technology, Big Pine CA 93513, USA.}

\author{Rick Hobbs}
\affiliation{Owens Valley Radio Observatory, California Institute of Technology, Big Pine CA 93513, USA.}

\author{David Hodge}
\affiliation{Cahill Center for Astronomy and Astrophysics, MC 249-17 California Institute of Technology, Pasadena CA 91125, USA.}

\author{Mark Hodges}
\affiliation{Owens Valley Radio Observatory, California Institute of Technology, Big Pine CA 93513, USA.}

\author{James W. Lamb}
\affiliation{Owens Valley Radio Observatory, California Institute of Technology, Big Pine CA 93513, USA.}

\author{Paul Rasmussen}
\affiliation{Owens Valley Radio Observatory, California Institute of Technology, Big Pine CA 93513, USA.}

\author{Jun Shi}
\affiliation{Cahill Center for Astronomy and Astrophysics, MC 249-17 California Institute of Technology, Pasadena CA 91125, USA.}

\author{Dana Simard}
\affiliation{Cahill Center for Astronomy and Astrophysics, MC 249-17 California Institute of Technology, Pasadena CA 91125, USA.}

\author{Jean Somalwar}
\affiliation{Cahill Center for Astronomy and Astrophysics, MC 249-17 California Institute of Technology, Pasadena CA 91125, USA.}

\author{Reynier Squillace}
\affiliation{Cahill Center for Astronomy and Astrophysics, MC 249-17 California Institute of Technology, Pasadena CA 91125, USA.}
\affiliation{Steward Observatory, University of Arizona, 933 N. Cherry Avenue, Tucson, AZ 85721, USA.}

\author{Sander Weinreb}
\affiliation{Cahill Center for Astronomy and Astrophysics, MC 249-17 California Institute of Technology, Pasadena CA 91125, USA.}

\author{David P. Woody}
\affiliation{Owens Valley Radio Observatory, California Institute of Technology, Big Pine CA 93513, USA.}

\author{Nitika Yadlapalli}
\affiliation{Cahill Center for Astronomy and Astrophysics, MC 249-17 California Institute of Technology, Pasadena CA 91125, USA.}

\collaboration{200}{(The Deep Synoptic Array team)}

\begin{abstract}

Faraday rotation measures (RMs) of fast radio bursts (FRBs) offer the prospect of directly measuring extragalactic magnetic fields. We present an analysis of the RMs of ten as yet non-repeating FRBs detected and localized to host galaxies by the 110-antenna Deep Synoptic Array (DSA-110). We combine this sample with published RMs of 15 localized FRBs, nine of which are repeating sources. For each FRB in the combined sample, we estimate the host-galaxy dispersion measure (DM) contributions and extragalactic RM. We find compelling evidence that the extragalactic components of FRB RMs are often dominated by contributions from the host-galaxy interstellar medium (ISM). Specifically, we find that both repeating and as yet non-repeating FRBs show a correlation between the host-DM and host-RM in the rest frame, and we find an anti-correlation between extragalactic RM (in the observer frame) and redshift for non-repeaters, as expected if the magnetized plasma is in the host galaxy. Important exceptions to the ISM origin include a dense, magnetized circum-burst medium in some repeating FRBs, and the intra-cluster medium (ICM) of host or intervening galaxy clusters. We find that the estimated ISM magnetic-field strengths, $\bar{B}_{||}$, are characteristically larger than those inferred from Galactic radio pulsars. This suggests either increased ISM magnetization in FRB hosts in comparison with the Milky Way, or that FRBs preferentially reside in regions of increased magnetic-field strength within their hosts.

\end{abstract}

\keywords{Cosmic electrodynamics (318), Extragalactic magnetic fields (507), Radio transient sources (2008), Neutron stars (1108), Polarimetry (1278), Radio pulsars (1353), Pulsars (1306)}

\section{Motivation and background}\label{introduction}

The origins, evolution, and physical states of magnetic fields in galaxies remain poorly understood, despite their importance. During the processes of galaxy evolution, primordial seed fields are likely amplified through dynamo processes \citep{rees87,beck96araa,beck2016magnetic}, flux freezing in gas collapse \citep{marinacci2015} and fragmentation \citep{su18mag}, and turbulence \citep{ryu08turb}. In the Milky Way and nearby spiral galaxies, magnetic fields are dynamically important in the interstellar medium (ISM), with comparable energy density to turbulent gas motions and cosmic rays, and provide significant pressure support \citep{beck2016magnetic}. Magnetism is critical to the processes of star formation in molecular clouds \citep{crutcher12}, and the propagation of cosmic rays and the resulting heating of the ISM and galaxy-scale winds \citep{zweibel13}. Estimates of the ISM magnetic-field strengths range from $\sim5$\,$\mu$G in the Solar neighborhood \citep[from direct Voyager-1 measurements;][]{v1}, to $\sim10$\,$\mu$G in denser regions of the Galactic plane \citep{crutcher12}, to several tens of $\mu$G in the ISM of nearby starburst galaxies \citep{beck2016magnetic}. Large-scale ordered fields are typically $\lesssim30\%$ of the total field strengths \citep{haverkorn2015,beck2016magnetic}. Simulations suggest that the large-scale fields, likely found mostly in disk-dominated galaxies \citep{marinaci2018}, saturate at several $\mu$G within a few Gyr after disk formation \citep{su18mag,rodrigues2019evolution}. Current estimates of magnetic-field strengths in external galaxies largely rely on equipartition analyses of synchrotron emission\footnote{Although see \citet{robishaw08} for observations of Zeeman splitting in OH megamasers.}, which are subject to significant uncertainties in cosmic-ray energy losses \citep[e.g.,][]{ponnada2022}. Direct measurements of magnetic-field strengths in external galaxies are required to test our understanding of the development of cosmic magnetism. 

Fast radio bursts (FRBs) offer a direct probe of extragalactic magnetic fields through combined measurements of Faraday rotation measures (RMs) and dispersion measures (DMs) \citep[e.g.,][]{masui2015,ravi2016magnetic,mannings2022fast}.  RMs and DMs together enable the line-of-sight integrated magnetic field ($\bar{B}_{||}$) to be evaluated (Note the $\bar{}$ indicates this is averaged over the line-of-sight). Independent estimates of the Galactic RM and DM foreground can be used to identify extragalactic contributions, and host-galaxy components can be further isolated using models for the DM of the intergalactic medium \citep[IGM;][]{Macquart_2020}. In this work, we assume that FRB RMs and DMs include contributions as follows:
\begin{eqnarray}
{\rm RM} &=& {\rm RM}_{\rm ion} + {\rm RM}_{\rm MW} + {\rm RM}_{\rm IGM} + \frac{{\rm RM}_{\rm host}}{(1 + z)^2} \\
\mathrm{DM} &=& \mathrm{DM}_{\rm MW} + \mathrm{DM}_{\rm MW,halo} + \mathrm{DM}_{\rm IGM}(z) + \frac{\mathrm{DM}_{\rm host}}{(1 + z)}.
\end{eqnarray}
Here, `ion' refers to the ionosphere contribution \citep{sotomayor2013ionfr}, `MW' refers to the Milky Way \citep[estimated for RMs and DMs respectively using:][]{hutschenreuter2022galactic,cordes2002ne2001}, and `host' refers to the FRB host galaxy. We assume a nominal Milky Way halo DM contribution of 10\,pc\,cm$^{-3}$ \citep[][our results are largely insensitive to this assumption]{ravi2023deep}, and absorb host-halo RM and DM contributions in the host-galaxy terms. We further define ${\rm RM}_{\rm exgal}\equiv{\rm RM}_{\rm IGM} + \frac{{\rm RM}_{\rm host}}{(1 + z)^2}$, and
\begin{eqnarray}
    \bar{B}_{||} &=& \frac{\mathrm{RM}}{0.81 \cdot \mathrm{DM}}\,\mu{\rm G} \\
    \bar{B}_{||,\,{\rm host}} &=& \frac{\mathrm{RM}_{\rm host}}{0.81 \cdot \mathrm{DM}_{\rm host}}\,\mu{\rm G}.
\end{eqnarray}

At present, 38 FRB sources have published polarization and/or RM data, of which just 15 (including six as yet non-repeating events) are localized to host galaxies (see Appendix~A of the companion paper by Sherman et al. for a detailed compilation). The RMs of FRBs span an extraordinary range, 
both in magnitude and variability in the case of repeating sources. FRB\,20121102A, for example, has an RM that varied from $1.3\times10^{5}$\,rad\,m$^{-2}$ to $7\times10^{4}$\,rad\,m$^{-2}$ \citep{michilli2018dynamic, plavin2022frb} over three years, which with a modestly high host DM \citep[around $50-225$\,pc\,cm$^{-3}$;][]{tendulkar2017} indicates a dense, dynamic magnetoionic environment. Similar environments are inferred for a few repeating-FRB sources \citep{anna2022highly,xu2022complex,mckinven2023revealing}, whereas others exist in entirely unremarkable magnetic environments \citep{kirsten2022glob,2023arXiv230414671F}. Frequency-dependent depolarization has also been observed in multiple repeating FRBs \citep{feng2022frequency,anna2022highly,mckinven2023revealing}, which may be described by the stochastic-RM model \citep{melrose1998stochastic, beniamini2022faraday, 2022ApJ...928L..16Y}. A polarization analysis by \citet{mannings2022fast} of five as yet non-repeating FRBs with host galaxy localizations could not disentangle any significant local RM contributions from broader host-ISM contributions. However, a tentative correlation between $\mathrm{DM}_{\rm host}$ and $\mathrm{RM}_{\rm host}$ suggested a significant contribution from ISM magnetic fields to the RMs of non-repeating FRBs.  

Here we augment the existing sample of RMs for 15 FRBs localized to host galaxies with ten as yet non-repeating FRBs from the 110-antenna Deep Synoptic Array (DSA-110) with significant RM detections and host galaxy localizations. Polarimetry and RM estimation for these new events were presented in the companion paper by Sherman et al., and host-galaxy analysis was presented in \citet{2023arXiv230703344L}. Our goals in this paper are to identify extragalactic and host-galaxy contributions to FRB RMs, and to estimate and interpret $\bar{B}_{||,\,{\rm host}}$ along the FRB sightlines. In \S\ref{sec:host} we outline our methods for disentangling various RM and DM contributions. We then discuss the origins of FRB RMs in \S\ref{sec:RMs}, and present and interpret measurements of $\bar{B}_{||,\,{\rm host}}$ in \S\ref{sec:Bp}. We discuss limitations of our work and future prospects in \S\ref{sec:disc}, and conclude in \S\ref{sec:summ}. Throughout we adopt cosmological parameters from the \textit{Planck} mission \citep{refId0}. RM and polarization data for the DSA-110 FRBs can be accessed through the DSA-110 Archive\footnote{\url{https://code.deepsynoptic.org/dsa110-archive/}}, with additional data made available upon request to the corresponding author \citep{Morrell_DSA-110_Event_Archive_2022}.

\section{Estimation of host DM and RM}\label{sec:host}

For each of the 10 FRB sources in the sample under consideration, we estimate DM$_{\rm host}$ following the methods outlined in \citet{connor2023dsa} \citep[see also][]{yang2022finding}. We derive the probability density function (PDF) of DM$_{\rm host}$ using the convolution of the PDFs of the observed DM with each DM component, as described in \citet{connor2023dsa}. $\mathrm{DM}_{\rm MW}$ is taken to have a Gaussian distribution with standard deviation 30\,pc\,cm$^{-3}$, $\mathrm{DM}_{\rm MW,halo}$ is taken to have a uniform distribution between 0--20\,pc\,cm$^{-3}$, and $\mathrm{DM}_{\rm IGM}$ is distributed according to \citet{zhang2021intergalactic} for each redshift. The distribution of $\rm DM_{\rm host}$ (in the observer frame) is then given by the convolution below:

\begin{equation}
P\left(\frac{\rm DM_{\rm host}}{1+z}\right) = P(\mathrm{DM})\ast P(-\mathrm{DM}_{\mathrm{MW}}) \ast P(-\mathrm{DM}_{\rm MW,halo}) \ast P(-\mathrm{DM}_{\rm IGM}). 
\end{equation}
The quoted values of DM$_{\rm host}$ in Table~\ref{table:RMTable} represent the expected values according to the PDFs.

We derive estimates of RM$_{\rm host}$ using a similar technique. The ionospheric contributions are estimated using the \textit{ionFR} Python library, which uses ionospheric data from NASA's Archive of Space Geodesy Data\footnote{\url{https://cddis.nasa.gov}/} to estimate RM for a given time and line of sight \citep{sotomayor2013ionfr}. The module was modified slightly to account for the change in format of NASA's data files since 2014 to use 1-hr timesteps. In general, ionospheric RM is of order $1$\,rad\,m$^{-2}$, but can change drastically in the presence of strong solar activity. The Galactic contribution is estimated using the \cite{hutschenreuter2022galactic} updated RM sky map.  While the Galactic contribution is more significant than ionospheric contributions, it remains small and is typically of order $10-20$\,rad\,m$^{-2}$. In deriving RM$_{\rm host}$, we neglect ${\rm RM}_{\rm IGM}$; this is justified below in Section~\ref{sec:RMs}. Table~\ref{table:RMTable} summarizes estimates of RM and DM components for each FRB in the DSA-110 sample. Note that $\rm DM_{host}$ (and by extension $\bar{B}_{||}$) could not be confidently constrained for FRB\,20220319D due to its close proximity to the Milky Way, as discussed in \citet{ravi2023deep}. $\rm RM_{host}$ and $\rm DM_{host}$ estimates are obtained for 9 repeating FRBs and 6 non-repeating FRBs from the published literature sample.

This analysis has several caveats. For example, host or intervening galaxy clusters can significantly affect some FRB DMs and RMs \citep{connor2023dsa,2023arXiv230605403L,ramesh2023azimuthal}, increasing the magnitudes of the estimated DM$_{\rm host}$ and RM$_{\rm host}$. Some repeating FRBs such as FRB\,20121102A and FRB\,20190520B show significant variation in RM \citep{anna2022highly,plavin2022frb}. For these sources, a weighted mean value for the RM is used, as the local contribution cannot be confidently removed. We also note that estimates of DM$_{\rm host}$ can be refined through a consideration of host-galaxy observables such as H$\alpha$ or UV luminosity \citep[e.g.,][]{mannings2022fast}, as well as observations of intervening galaxies \citep[e.g.,][]{2023arXiv230307387S}, but we defer such an analysis to future work. 

\begin{deluxetable*}{ccccccccc}
\tabletypesize{\scriptsize}
\tablewidth{0pt} 
\tablecaption{DSA-110 FRB Sample Rotation Measure Properties\label{table:RMTable}}
\tablehead{\colhead{\textbf{FRB}} & \colhead{\textbf{Redshift}} & \colhead{\textbf{DM}} & \colhead{${\rm \mathbf{DM}}_{\rm \mathbf{host}}$} & \colhead{\textbf{RM}} & \colhead{${\rm \mathbf{RM}}_{\rm \mathbf{MW}}$} & \colhead{${\rm \mathbf{RM}}_{\rm \mathbf{ion}}$}  & \colhead{${\rm \mathbf{RM}}_{\rm \mathbf{host}}$}  & \colhead{$\mathbf{\bar{B}_{\rm \mathbf{||,host}}}$ ($\mathbf{\mu}$\textbf{G})}}
\startdata 
20220207C &  0.043040 &                                            $262.3$ &                     $136.98^{+ 39.31 }_{- 38.72 }$ &                                 $162.48 \pm 0.04 $ &                                 $-5.22 \pm 16.93 $ &                                   $1.35 \pm 0.07 $ &                                $181.28 \pm 18.42 $ &                       $1.97^{+ 0.16 }_{ - 0.84 } $ \\
\hline
20220307B & 0.248123 &                                           $499.15$ &                     $169.77^{+ 76.15 }_{- 83.96 }$ &                               $-947.23 \pm 12.27 $ &                                 $-4.11 \pm 28.64 $ &                                   $1.85 \pm 0.09 $ &                               $-1477.8 \pm 48.74 $ &                      $-13.16^{+ 5.72 }_{ - 5.3 } $ \\
\hline
20220310F  &  0.477958 &                                           $462.15$ &                       $66.5^{+ 49.16 }_{- 48.46 }$ &                                  $11.39 \pm 0.19 $ &                                 $-14.33 \pm 7.43 $ &                                   $0.51 \pm 0.09 $ &                                  $55.9 \pm 16.26 $ &                         $2.64^{+ 0.9 }_{ - 2.1 } $ \\
\hline
20220319D & 0.011228 &                                           $110.95$ &                                                 -- &                                 $59.94 \pm 14.33 $ &                                $-13.82 \pm 17.68 $ &                                   $2.07 \pm 0.09 $ &                                 $73.63 \pm 23.26 $ &                                                 -- \\
\hline
20220418A &  0.622000 &                                           $623.45$ &                     $109.04^{+ 76.21 }_{- 76.92 }$ &                                 $6.13 \pm 7.48$ &                                  $7.96 \pm 13.96 $ &                                    $0.63 \pm 0.1 $ &                                 $-5.76 \pm 41.67 $ &                         $-0.17^{+ 0.61 }_{ - 0.60 } $ \\
\hline
20220506D & 0.30039 &                                           $396.93$ &                      $90.63^{+ 56.86 }_{- 59.16 }$ &                                  $-32.38 \pm 3.6 $ &                                 $-9.68 \pm 14.72 $ &                                   $1.16 \pm 0.08 $ &                                $-39.75 \pm 25.62 $ &                      $-1.27^{+ 1.03 }_{ - 0.38 } $ \\
\hline
20220509G & 0.089400 &                                            $269.5$ &                     $129.16^{+ 47.13 }_{- 49.62 }$ &                                 $-109.0 \pm 1.17 $ &                                 $14.03 \pm 14.05 $ &                                   $0.69 \pm 0.08 $ &                               $-146.54 \pm 16.73 $ &                      $-1.98^{+ 0.88 }_{ - 0.33 } $ \\
\hline
20220825A & 0.241397 &                                            $651.2$ &                   $395.77^{+ 112.01 }_{- 123.38 }$ &                                 $750.23 \pm 6.67 $ &                                 $-3.53 \pm 15.26 $ &                                   $0.32 \pm 0.03 $ &                               $1161.38 \pm 25.67 $ &                       $4.49^{+ 0.63 }_{ - 1.77 } $ \\
\hline
20220920A & 0.158239 &                                            $315.0$ &                     $136.33^{+ 57.15 }_{- 62.41 }$ &                                $-830.25 \pm 8.29 $ &                                  $0.72 \pm 13.12 $ &                                    $1.6 \pm 0.09 $ &                              $-1117.08 \pm 20.83 $ &                     $-12.29^{+ 5.08 }_{ - 4.34 } $ \\
\hline
20221012A &  0.284669 &                                            $442.2$ &                     $162.95^{+ 78.21 }_{- 85.64 }$ &                                 $165.7 \pm 17.66 $ &                                  $3.07 \pm 12.45 $ &                                   $1.69 \pm 0.13 $ &                                $265.86 \pm 35.62 $ &                       $3.27^{+ 0.65 }_{ - 1.96 } $ \\
\hline
\enddata
\bigskip
\textit{Notes:} All DMs are expressed in units of pc\,cm$^{-3}$, and all RMs are expressed in units of rad\,m$^{-2}$.
\end{deluxetable*}


\section{The origin of FRB rotation measures}\label{sec:RMs}

\begin{figure}
    \centering
    \includegraphics[width=0.9\textwidth]{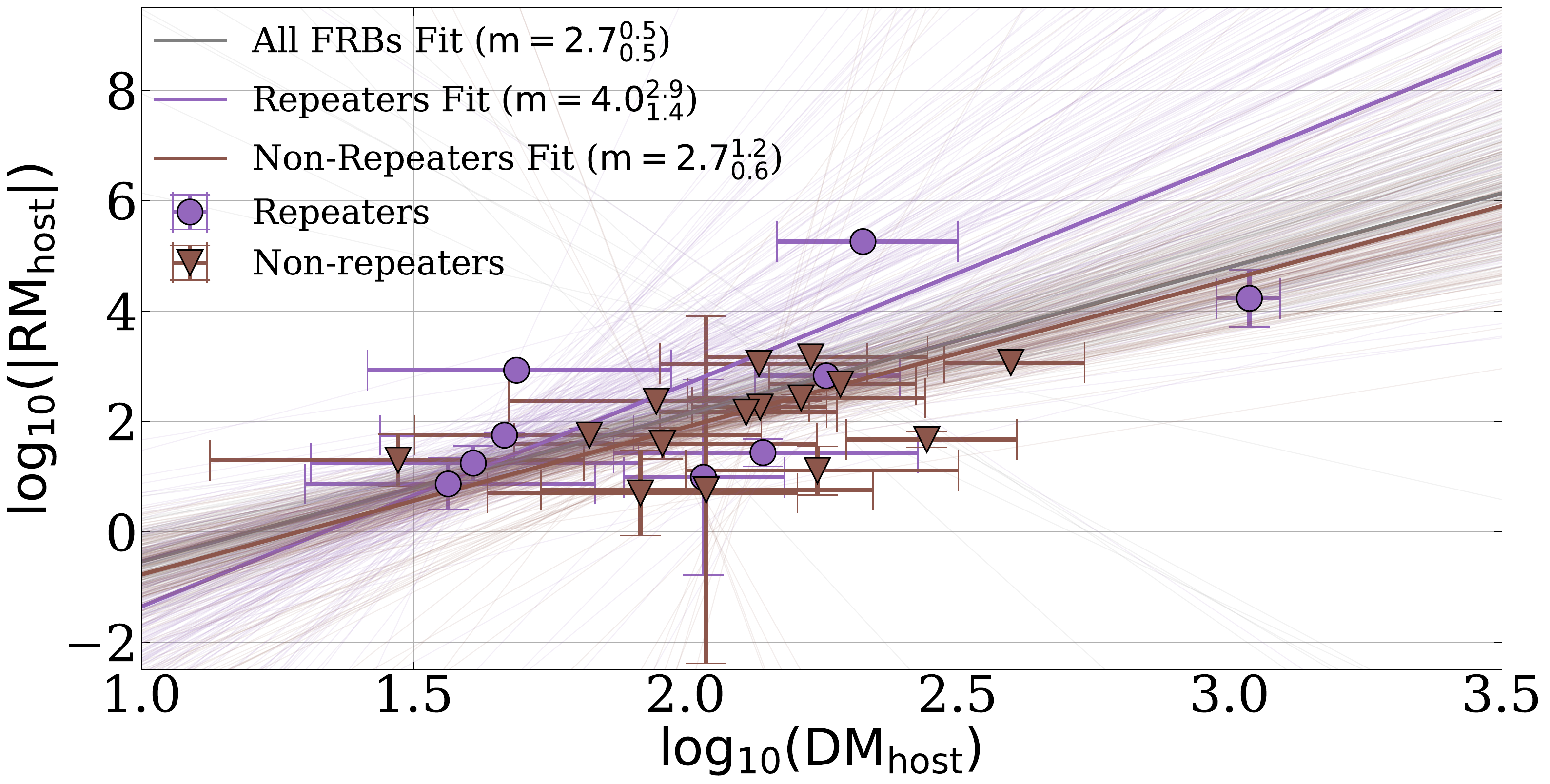}
    \caption{Estimates of ${\rm RM}_{\rm host}$ and ${\rm DM}_{\rm host}$ for as yet non-repeating (brown triangles) and repeating FRBs (purple circles). Solid lines show the best fit from a Markov-Chain Monte Carlo (MCMC) analysis for non-repeaters (brown), repeaters (purple), and the full sample (grey). Random draws from each MCMC simulation are shown by faint lines for non-repeaters (brown), repeaters (purple) and all FRBs (grey). The best fit slopes with 1-$\sigma$ errors are displayed in the legend in each case. }
    \label{fig:rmdm}
\end{figure}

\begin{figure}
    \centering
    \includegraphics[width=0.9\textwidth]{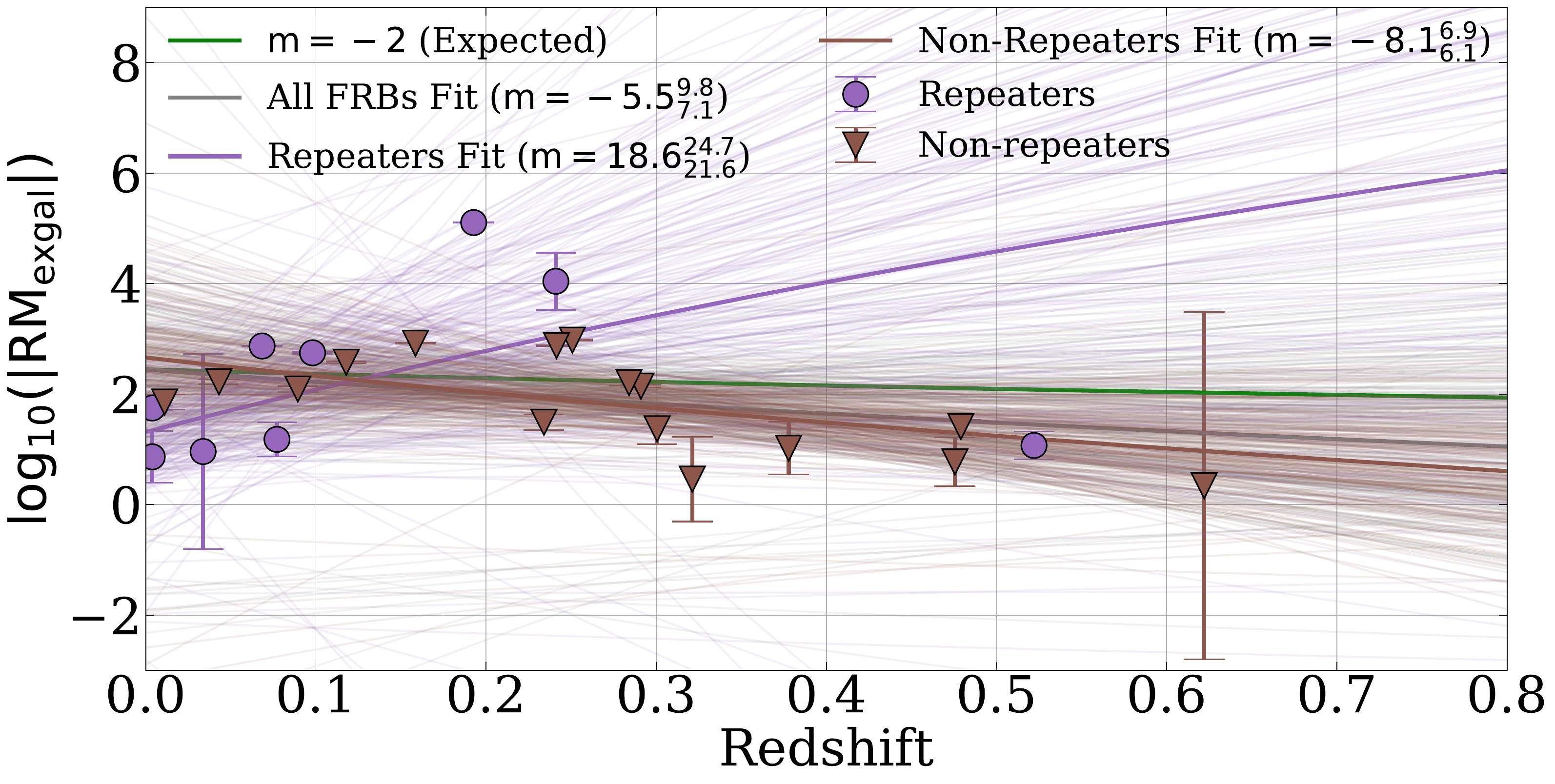}
    \caption{Estimates of ${\rm RM}_{\rm exgal}$ at different redshifts for as yet non-repeating (brown triangles) and repeating FRBs (purple circles). Solid lines show the best fit from an MCMC analysis for non-repeaters (brown), repeaters (purple), and the full sample (grey). The green line indicates the expected scaling from redshift suppression, i.e. ${\rm RM}_{\rm exgal} \propto (1+z)^{-2}$. Samples from each MCMC simulation are shown by faint lines for non-repeaters (brown), repeaters (purple) and all FRBs (grey). The best fit slopes with 1-$\sigma$ errors are displayed in the legend in each case. Significant negative correlation is found for non-repeaters ($p=0.006$) using a Spearman Rho test, although the median slope $m = -8.1^{6.9}_{6.1}$ is only marginally consistent with redshift suppression. The two repeaters with the highest values of ${\rm RM}_{\rm exgal}$, FRB\,20121102A and FRB\,20190520B, are known to have highly magnetized, dynamic local environments that may dominate their RMs \citep{anna2022highly,plavin2022frb}.}
    \label{fig:rmz}
\end{figure}

We first test the hypothesis that the host-galaxy ISM predominantly determines FRB RMs. We do this by searching for a correlation between ${\rm RM}_{\rm host}$ and ${\rm DM}_{\rm host}$. Tentative evidence ($p=0.06$, where $p$ is the probability of the null hypothesis of no correlation) for such a correlation was identified by \citet{mannings2022fast} using just nine FRBs (three repeaters and six non-repeaters). When considering all extragalactic contributions to the RMs and DMs of FRBs, such a correlation is the best motivated, both empirically and theoretically. Large-scale fields in the IGM are not evident in studies of the RMs of active galactic nuclei \citep{2012arXiv1209.1438H}. As such fields are likely no larger than several nano-Gauss, they will not contribute more than a few units of RM for typical FRBs \citep[e.g.,][]{2016ApJ...824..105A,ravi2016magnetic,2019MNRAS.488.4220H}. Furthermore, both observation of nearby galaxies \citep{heesen2023detection} and IllustrisTNG simulations \citep{ramesh2023azimuthal} imply the circum-galactic media (CGM) of intervening galaxies contribute negligibly to the observed RM ($\sim 0.01$\,rad\,m$^2$ at large impact parameters $\gtrsim200$\,kpc) compared to host ISM, local, and Milky Way contributions. Galaxy clusters, which will contribute significantly more RM \citep{connor2023dsa}, likely affect on the order of 10\% of FRB sightlines and may not dominate the IGM DM contributions. On the other hand, experience from Galactic pulsar observations suggests that pulsar RMs are excellent tracers of large-scale components of the Galactic magnetic field via correlations with DM \citep{han2018mag,sobey19}. In the case of FRBs (and pulsars), a correlation between ${\rm RM}_{\rm host}$ and ${\rm DM}_{\rm host}$ can be masked if FRBs generally occupy dynamic local environments that contribute large RMs but little DM, like the environment of FRB\,20121102A, because the DM component would be small with respect to other sources of DM.

We have conducted simple, physically motivated Monte Carlo simulations to demonstrate how extreme (either dynamic or highly magnetized) local FRB environments could mask the correlation between $\rm RM_{\rm host}$ and $\rm DM_{\rm host}$. For a population of 24 FRBs, we draw samples of $\rm RM_{\rm ISM,host}$ and $\rm DM_{\rm ISM,host}$ assuming they have similar distributions to those of Milky Way pulsars. A subset of FRBs are then randomly selected to have extreme local environments; we consider cases in which the local magnetic field is (1) stochastically varying with Root Mean Squared (RMS) $\sigma_{\bar{B}_{||},\rm local} = 100 \, \mu$G, (2) smoothly varying with rate of change $d{\bar{B}_{||,\rm local}}/dt = 0.1\,\mu$G\,day$^{-1}$, or (3) constant and of extreme magnitude $|\bar{B}_{||,\rm local}| =100\, \mu $G\footnote{These nominal values are estimated to match the environments of FRBs with known dynamic local environments such as FRB\,20121102A and FRB\,20190520B \citep[e.g.,][]{mckinven2023revealing}.}. The local DM is drawn from a uniform distribution between $0-10\,$pc\,cm$^{-3}$, and the local RM contribution computed as $\rm RM_{\rm local} \approx 0.81\bar{B}_{\rm ||,local}DM_{\rm local}$. The observed  $\rm RM_{\rm host}$ and $\rm DM_{\rm host}$  are computed from the sum of the local and ISM contributions. We find that if the local field is either stochastically or smoothly varying, the correlation (for a Spearman Rho test with a 90\% confidence level) between $\rm RM_{\rm host}$ and $\rm DM_{\rm host}$ can be masked if $\gtrsim 30-55\%$ of the sample ($\gtrsim 8-13$ FRBs) have extreme local environments, depending on the mean value of $\bar{B}_{||,\rm local}$. If the local field is constant, the correlation can be masked if only $\gtrsim 8\%$ of the sample ($\gtrsim 7-8$ FRBs) have $|\bar{B}_{||,\rm local}| =100\,\mu $G. Therefore, if a correlation is observed, one can conclude that the host ISM dominates the observed $\rm RM_{\rm host}$ and $\rm DM_{\rm host}$, while an upper limit $\lesssim 55\%$) of FRBs may reside in extreme local environments.


Figure~\ref{fig:rmdm} shows a clearly detected correlation between ${\rm RM}_{\rm host}$ and ${\rm DM}_{\rm host}$. Significant correlations are found for repeaters ($p=0.02$), non-repeaters ($p=0.06$), and for all FRBs ($p=0.004$) using Spearman Rho tests. This confirms the tentative result of \citet{mannings2022fast}, and shows that it applies to both repeating and non-repeating FRBs. The existence of this correlation indicates that the RMs of both repeaters and non-repeaters are generally determined by large-scale fields in host-galaxy ISM. The ISM origin is bolstered by the weak correlation between host RM/DM and galactocentric offset found by \citet{mannings2022fast}, which ought not to exist if the magnetized plasma were in circumburst material.

An analysis of observed ${\rm RM}_{\rm exgal}$ evolution with redshift $z$ also demonstrates that RM originates in the host galaxy. As shown in Figure~\ref{fig:rmz}, we find a significant negative correlation ($p=0.006$) between ${\rm RM}_{\rm exgal}$ and $z$ for non-repeaters. The result for repeating FRBs is significantly affected by FRB\,20121102A and FRB\,20190520B, which are known to have highly magnetized, dynamic local environments that may dominate their RMs \citep{anna2022highly,plavin2022frb}. However, this does not contradict the conclusion that the host ISM dominates the RM and DM of repeaters. If repeating FRBs have higher variance in RM (both between sources and from burst-to-burst), then that variance could drown out the relatively weak $(1+z)^{-2}$ effect. The lack of repeating FRBs at high redshift ($z > 0.3$) may also explain why local environments obscure the redshift suppression, but a significant correlation is still found between ${\rm RM}_{\rm host}$ and ${\rm DM}_{\rm host}$. For non-repeaters, a negative correlation between ${\rm RM}_{\rm exgal}$ and $z$ can most simply be attributed to the suppression of a characteristic ${\rm RM}_{\rm host}$ by a $(1+z)^{-2}$ term, although this possibility is only marginally consistent with the measured slope of the anti-correlation. Cosmic evolution in host-galaxy field strengths may also play a role, although there is no theoretical basis for the observed ${\rm RM}_{\rm exgal}$-$z$ anti-correlation \citep[e.g.,][]{rodrigues2019evolution}.

\section{Magnetic fields in FRB environments and host galaxies}\label{sec:Bp}

\begin{figure}
    \centering
    \includegraphics[width=0.9\textwidth]{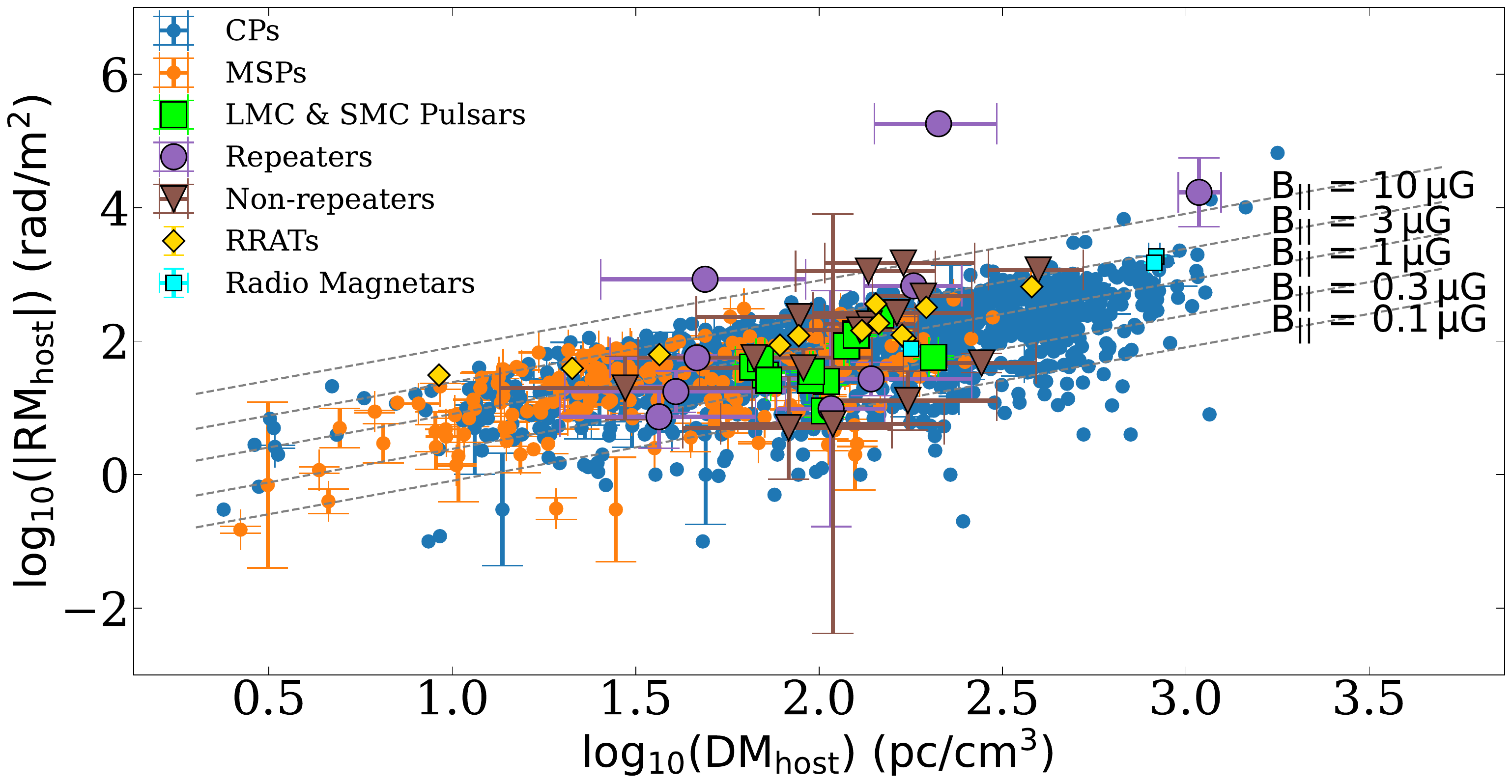}
    \caption{Comparison of FRB ${\rm RM}_{\rm host}$ and ${\rm DM}_{\rm host}$ estimates with the RMs and DMs of pulsars, RRATs, and radio-loud magnetars. As yet non-repeating  (brown triangles) and repeating FRBs (purple circles) are distinguished, as are canonical pulsars (CPs; $P > 30\, $ms; blue circles) and millisecond pulsars (MSPs; $P < 30 \, $ms; orange circles). RRATs (yellow diamonds), pulsars in the Large and Small Magellanic Clouds (LMC, SMC; green squares), and radio magnetars (teal squares) are included to complete the sample. Dotted grey lines of constant magnetic field, $\bar{B}_{\rm ||,host}$, are shown. Both repeaters and non-repeaters are found to have distinct $\bar{B}_{\rm ||,host}$ from CPs and MSPs using a Kolmogorov-Smirnov (K-S) test and an Anderson-Darling (A-D) test with $>99\%$ confidence. FRBs in general appear to occupy slightly more magnetized environments than pulsars. K-S and A-D tests find similarity between FRB and RRAT distributions. Data on literature samples of neutron stars were compiled according to Appendix~A of the companion paper by Sherman et al.}
    \label{fig:psr}
\end{figure}

\begin{figure}
    \centering
    \includegraphics[width=0.9\textwidth]{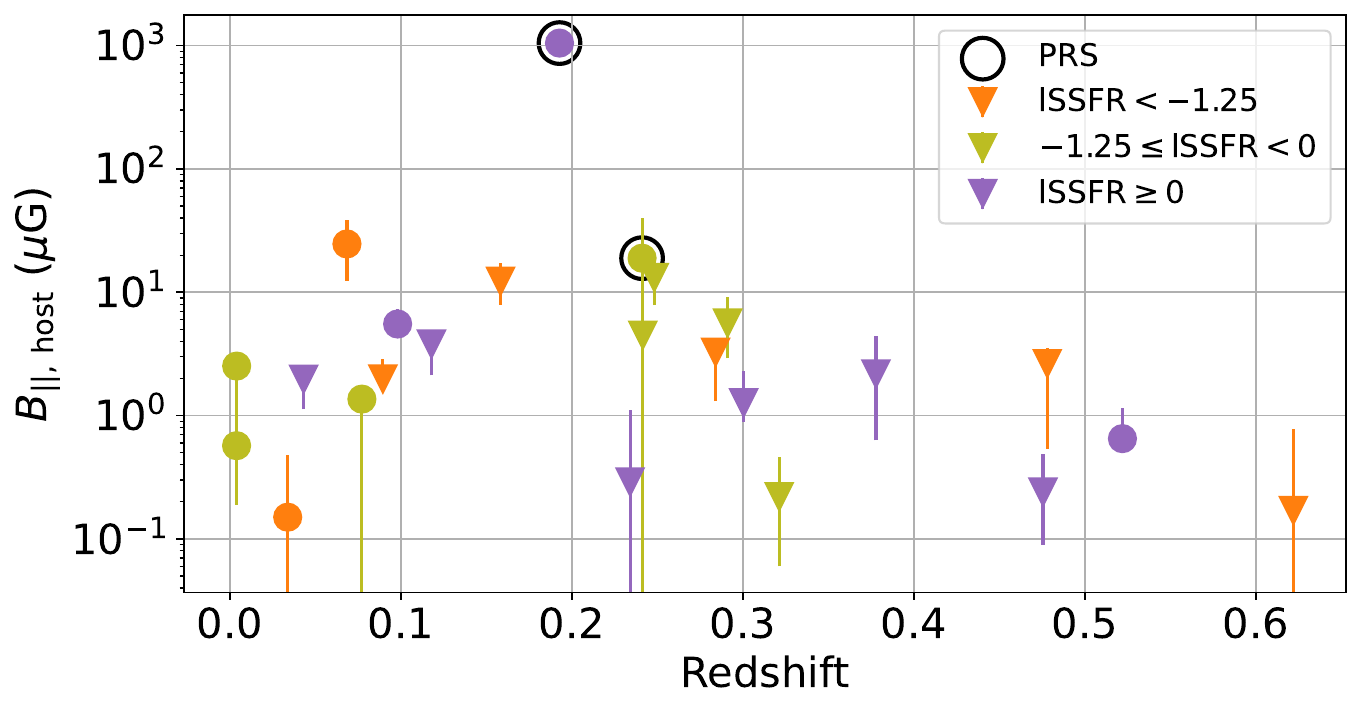}
    \caption{Estimates of $\bar{B}_{\rm ||,host}$ for FRBs at different redshifts, including as yet non-repeating sources (triangles) and repeaters (circles). FRBs are grouped by the (logarithmic) specific star-formation rates (lSSFR) of the host galaxies, scaled by $10^{10}M_{\odot}$: $\rm lSSFR < -1.25$ (orange), $\rm -1.25 \le lSSFR < 0$ (green) and $\rm lSSFR \ge 0$ (purple). FRBs 20121102A and 20190520B, both associated with persistent radio sources \citep[PRSs;][]{2022ApJ...927...55L}, are circled in black. No correlation is found between $\bar{B}_{\rm ||,host}$ and redshift for any subgroup, nor are there significant differences between the subgroups. }
    \label{fig:bz}
\end{figure}

Having established that estimates of ${\rm RM}_{\rm host}$ for our FRB sample may trace the large-scale magnetic fields in the host-galaxy ISM, we can estimate and interpret $\bar{B}_{\rm ||,host}$ for the sample. The estimates for the DSA-110 sample are given in Table~\ref{table:RMTable}. We focus first on comparing estimates of $\bar{B}_{\rm ||,host}$ with estimates of $\bar{B}_{\rm ||}$ for pulsars. This analysis is complementary to the investigation by \citet{chrimesPSR} of the physical locations of FRBs in their hosts relative to pulsars in the Milky Way. It is, however, similarly motivated by the prospect of determining whether or not Galactic pulsars are similarly located within the Milky Way as FRBs are within their hosts. 

Figure~\ref{fig:psr} shows the distribution of ${\rm RM}_{\rm host}$ and ${\rm DM}_{\rm host}$ for FRBs in comparison with RMs and DMs of Galactic and Magellanic-Cloud pulsars, rotating radio transients (RRATs), and radio-loud magnetars. We find that the distributions of $\bar{B}_{\rm ||,host}$ for repeating and as yet non-repeating FRBs are significantly different to the $\bar{B}_{\rm ||}$ distributions for pulsars and magnetars ($p \le 0.089$ for repeaters; $p \le 0.006$ for nonrepeaters when compared to CPs, MSPs, and all pulsars/magnetars), while the comparison to RRATs is inconclusive. Specifically, FRBs tend to have higher inferred values of  $\bar{B}_{\rm ||,host}$ than Galactic neutron stars. There are two leading possibilities for this result. First, a significant fraction of FRBs, both repeating and non-repeating, may have important RM contributions from circum-source environments that contribute negligibly to their DMs \citep[e.g.,][]{2023MNRAS.520.2039Y}. Such RM contributions would need to be significant enough to raise the inferred $\bar{B}_{\rm ||,host}$ values by some tens of percent, while not masking the ${\rm RM}_{\rm host}$--${\rm DM}_{\rm host}$ correlation\footnote{Note some outlier repeating FRBs such as FRB\,20121102 and FRB\,20190520B have less stringent requirements for significant RM given their already high inferred $\bar{B}_{\rm ||,host}$} . Alternatively, FRBs may be typically hosted by galaxies with larger large-scale ISM field strengths than observed in the Milky Way. This idea is consistent with observational and theoretical work establishing an increase in large-scale fields at fixed stellar mass with redshift, and the variation in field strength with star-formation activity \citep{beck2016magnetic,rodrigues2019evolution}.

We can test the second scenario by investigating whether either redshift or specific star-formation rate (sSFR) are correlates of $\bar{B}_{\rm ||,host}$ (Figure~\ref{fig:bz}). Estimates of sSFR for the literature sample of localized FRBs are derived from \citet{gordon2023} where possible, and otherwise from \citet{bhardwaj2021nearby}, \citet{michilli2022subarcminute}, and \citet{ravi2022deep}. Estimates of sSFR for DSA-110 FRBs are sourced from \citet{2023arXiv230703344L}. We find no significant correlation between $\bar{B}_{\rm ||,host}$ and either redshift or sSFR, either for repeaters or non-repeaters. This result disfavors cosmic evolution and a biased host-galaxy population as explanations for the characteristically higher $\bar{B}_{\rm ||,host}$ values of FRBs in comparison with Galactic neutron stars. However, we cannot immediately favor the importance of circum-source environments in determining $\bar{B}_{\rm ||,host}$. First, we are not correcting for the location of the FRBs within their hosts, nor for the host-galaxy orientations, which may be also be important in determining FRB RMs \citep[e.g.,][]{kirsten2022glob,mannings2022fast}. Second, as is evident from Figure~\ref{fig:psr}, it is possible that only some FRBs have characteristically higher $\bar{B}_{\rm ||,host}$ than Galactic pulsars. These objects, with $\bar{B}_{\rm ||,host}\gtrsim10$\,$\mu$G, likely do have significant circum-source RM contributions, additionally because large-scale fields are expected to saturate below 10\,$\mu$G in most galaxies.

\section{Discussion}\label{sec:disc}

Unlike with DM, the IGM and circum-galactic medium (CGM) of intervening galaxies are not expected to contribute significantly to the extragalactic RM of FRBs, due to small magnetic fields in those media \citep{2012arXiv1209.1438H, CGMRM2020, ponnada2022}. For example, \citet{IGMRM2021} found that the 3-$\sigma$ upper limit on the RM from filaments in the IGM is $<$\,3.8$\,\rm rad\,m^{-2}$, which is comparable to the Faraday column of Earth’s ionosphere. Instead, the extragalactic RM of FRBs will typically be dominated by the host galaxy. For $\mathcal{O}(10\%)$ of sources, the magnetized ICM of galaxy clusters will contribute non-negligibly to the observed RM. This can happen when the FRB sightline intersects a foreground galaxy cluster by chance \citep{190520cluster} as well as when the host galaxy is a cluster member \citep{connorcluster23, sharma2023massive}. DM and RM estimates for a sample of FRBs impacted by galaxy clusters will allow for magnetic-field constraints in the intracluster medium (ICM), including beyond the virial radius where probes such as thermal X-ray emission are less sensitive. 

The majority of FRB sightlines will \textit{not} be impacted by the ICM. For such sources, the component of extragalactic DM that correlates with RM$_{\rm exgal}$ ought to come from the host galaxy. Thus, for a large sample of localized FRBs at low or moderate redshifts, the relationship between extragalactic DM and RM could be used to extract the distribution of DM$_{\rm host}$. The host galaxy DM distribution is currently poorly constrained, but has significant implications for FRB applications to cosmology as well as in our understanding of FRB host galaxies and progenitor environments. In this case, one must assume that the host-galaxy RM and DM originate from the same plasma. 

In general, strong prospects remain for using FRBs to constrain or measure magnetic-field strengths in a variety of environments, including the ISM, galaxy groups, and galaxy clusters. If a sample of $z\gtrsim1$ FRBs can be obtained, the suppression of ${\rm RM}_{\rm host}$ may enable us to cleanly consider RM contributions external to galaxies in analogy to the estimation of DM contributions in the CGM and IGM \citep[e.g.,][]{2022NatAs...6.1035C}. Fortunately, a high-redshift sample of FRB RMs is likely to be discovered with upcoming surveys \citep{2023MNRAS.521.4024C}, thanks to their large overall detection rate and increased sensitivity \citep[e.g.,][]{2019clrp.2020...28V, hallinanDSA2k}.
Sub-arcsecond FRB localizations \citep[e.g.,][]{2019clrp.2020...28V} will further assist in modeling and interpreting ISM RM contributions.

\section{Conclusions}\label{sec:summ}

We present a study of the RMs of 25 FRBs localized to host galaxies, including ten new measurements from the DSA-110. The localizations enable a focused consideration of host-galaxy contributions to FRB RMs, and the resulting implications. We conclude the following.
\begin{enumerate}

    \item The RMs of FRBs are predominantly contributed by the host-galaxy ISM. This is evidenced by an observed correlation between ${\rm RM}_{\rm host}$ and ${\rm DM}_{\rm host}$ for both repeating and as yet non-repeating FRBs (Figure~\ref{fig:rmdm}), and supported by an anti-correlation between ${\rm RM}_{\rm exgal}$ and $z$ for non-repeating FRBs that suggests a characteristic value of ${\rm RM}_{\rm host}$ (Figure~\ref{fig:rmz}). Important known exceptions include FRBs hosted by or viewed through galaxy clusters \citep[e.g, FRB\,20220509G;][]{connor2023dsa}, and FRBs with highly magnetized circum-source environments \citep[e.g, FRBs 20121102A and 20190520B;][]{anna2022highly,plavin2022frb}. 

    \item The distribution of inferred $\bar{B}_{\rm ||,host}$ values for FRBs is generally inconsistent with the $\bar{B}_{\rm ||}$ distribution for Galactic and Magellanic-Cloud pulsars (Figure~\ref{fig:psr}). Higher values of $\bar{B}_{\rm ||,host}$ are observed for FRBs in comparison with pulsars. 

    \item We test the possibility of the $\bar{B}_{\rm ||,host}$ excess being explained by a biased FRB host-galaxy population with stronger large-scale magnetic fields than the Milky Way. We find no evidence for this scenario (Figure~\ref{fig:bz}); this result does not however exclude the possibility. The locations of FRBs within their hosts, which we do not account for, may be important in interpreting host-ISM magnetic-field estimates. It also remains plausible that a significant subset of FRBs have substantially magnetized circum-source environments that contribute to but do not dominate estimates of $\bar{B}_{\rm ||,host}$. 

\end{enumerate}

\begin{acknowledgments}

The authors would like to thank Jim Cordes, Dongzi Li, Bing Zhang, Yuanhong Qu, Joel Weisberg, and Sam Ponnada for insightful and essential conversations on polarization theory and direction on the analysis conducted, as well as Paul Bellan and Yang Zhang for a comprehensive Plasma Physics course. We also thank Yi Feng, Dipanjan Mitra, Yuan-Pei Yang, Dylan Nelson, and Reshma Anna-Thomas for useful comments and recommendations on the early draft. The authors thank staff members of the Owens Valley Radio Observatory and the Caltech radio group, including Kristen Bernasconi, Stephanie Cha-Ramos, Sarah Harnach, Tom Klinefelter, Lori McGraw, Corey Posner, Andres Rizo, Michael Virgin, Scott White, and Thomas Zentmyer. Their tireless efforts were instrumental to the success of the DSA-110. The DSA-110 is supported by the National Science Foundation Mid-Scale Innovations Program in Astronomical Sciences (MSIP) under grant AST-1836018. 

\end{acknowledgments}

\bibliography{Refs}{}
\bibliographystyle{aasjournal}

\end{document}